\documentclass[aps,pra,twocolumn,showpacs,floatfix]{revtex4}
\usepackage{graphicx}
\usepackage{epsfig}
\begin{document}
\setlength{\unitlength}{1mm}
\bibliographystyle{unsrt}
\title{Pump-probe measurement of atomic parity violation in caesium

 with a precision of 2.6$\%$}
 \author{M. Lintz\footnote{\emph{Present address:} Observatoire de
la C\^ote d'Azur, BP 4229, F-06304 Nice; michel.lintz@obs-nice.fr},
J. Gu\'ena\footnote{\emph{Present address:} METAS, Swiss Federal
Office of Metrology, CH-3003 Bern-Wabern; jocelyne.guena@metas.ch},
and M.A. Bouchiat } \affiliation{ Laboratoire Kastler
Brossel\footnote{Laboratoire Kastler Brossel is a Unit\'e de
Recherche de l'Ecole Normale Sup\'erieure et de l'Universit\'e
Pierre et Marie Curie, associ\'ee au CNRS (UMR 8552).}
\\D\'epartement de Physique de
l'Ecole Normale Sup\'erieure,\\ 24 Rue Lhomond, F-75231  Paris Cedex
05, France }
 \date{September 7, 2006}
 \begin{abstract}
We present the atomic parity violation measurements made in Cs
vapour using a pump-probe scheme. After pulsed excitation of the
6S-7S forbidden transition in the presence of a longitudinal
electric field, a laser beam resonant with  one of the
 7S-6P transitions stimulates the 7S atom emission for a duration of 20~ns. The polarisation of the amplified probe beam is
 analysed. A seven-fold signature allows discrimination of the parity violating linear dichroism, and real-time calibration
 by a similar, known, parity conserving linear dichroism. The zero-field linear dichroism signal due to the magnetic dipole
  transition moment is observed for the first time, and used for in-situ determination of the electric field. The result,
  $ImE_{1}^{pv}= (-808\pm 21) \times 10^{-14} ea_{0}$, is in perfect agreement with the corresponding, more precise measurement
  obtained by the Boulder group. A transverse field configuration with large probe amplification could bring atomic parity violation measurements
 to the 0.1$\%$ accuracy level.
\end{abstract}

\maketitle

\section{Introduction}\label{sec:1}
\label{intro} In this paper we present the PV measurements we have
performed in Cs \cite{Ref17}, by making use of a probe beam
superposed with the beam that drives the 6S-7S forbidden transition.
After the probe beam has been amplified by stimulated emission of
the excited vapour, its polarisation is analysed so as to extract
the PV asymmetry, calibrated against a parity conserving signal.
This experimental scheme has the potential for high precision PV
measurements, since the PV asymmetry to be measured is an increasing
function of the excited atom number density. The quantity of metal
necessary for vapour cell operation is orders of magnitude smaller
than in an atomic beam experiment, and would be acceptable for an
experiment with the long lived isotope $^{135}$Cs.
\section{Motivations}\label{sec:2}
\label{sec:1} Since the emergence of the field \cite{Ref1}, atomic
parity violation (APV) measurements in heavy atoms have evolved and
diversified \cite{Ref2}. These difficult experiments are motivated
by the fact that APV measurements provide unique inputs for testing
the Standard Model (SM). Measurements at the pole of the Z$^{0}$
boson mass energy  have reached an impressive precision, but
testing, for instance, the variation of the value of the weak mixing
angle with energy requires data taken at different energy scales,
such as the recent measurement of the PV asymmetry in M$\o$ller
scattering \cite{Ref3}, and the APV measurements in caesium. The
corresponding momentum transfers are about 160 MeV/c and a few MeV/c
respectively, to be compared with 100 GeV/c at the Z$^{0}$ pole.
Besides the energy scale, the information extracted from a
determination of the Cs weak charge $Q_W$ is a test in the hadronic
sector of the SM, and a test different from those extracted, for
instance, from the inelastic scattering processes exploited in
\cite{Ref4}. In a model-independent analysis of the results, the
corresponding allowed regions, in the plane spanned by the
elementary weak charges of the u and d quarks, turn out to be nearly
orthogonal.
\\
Relevance to physics beyond the SM is one more motivation. APV
experiments are particularly sensitive, among the P-violating
processes, to those possibly mediated by particles with a mass in
the MeV range, while such a "light" boson could escape detection in
high energy experiments \cite{Ref5}. The U boson predicted by some
supersymmetric extensions of the SM is a candidate to explain,
through dark matter recombination, the astrophysical observation
that the bulge of the galaxy shows a significant emission at 511 keV
\cite{Ref6}. However, to account for the moderate intensity of the
emission line, one has to assume \cite{Ref5} that the axial coupling
of the U boson (mass $\approx$ 10 MeV) to electrons is very small.
Indeed the present agreement between APV measurements in Cs and the
SM \cite{Ref7} excludes such an axial coupling at the $10^{-6}$
level. Some of the Kaluza-Klein models of space-time also make
predictions of effects observable in APV without counterparts in
high-energy experiments \cite{Ref8}.
\\
The Boulder measurement \cite{Ref9}, with two 0.5$\%$ precision
measurements on two hyperfine (HF) components of the Cs forbidden
transition, is the only experiment that could achieve a precision
test of the standard model at very low energy. Controlling the
systematics \cite{Ref10} has been a difficult aspect of this
experiment, and a cross-check of this measurement would be highly
valuable, in regard of the implications of  APV experiments. The
experimental scheme developed in Paris and presented in this paper
combines the advantages {\it{i)}} of the "Stark" (electric field
enhanced excitation) experiments: very well defined signature,
removing the need for scanning the excitation wavelength to identify
the PV signal, and hence the risk associated with the
"wavelength-dependent analysis angle" of the optical rotation
experiments in heavier atoms; and {\it{ii)}} of those latter
"transmission" experiments: use of the probe beam allows to detect
the whole excited vapour column with high efficiency.
\\
Most importantly, our experimental scheme has only two features in
common with those of the Boulder experiment: the choice of the
$6S_{1/2}-7S_{1/2}$ forbidden transition  and the application of a
Stark field, both suggested in \cite{Ref1}. As a result the
systematic effects are very different, a crucial point since the
measurements presented here aim at a cross-check of the Boulder
experiment. Some of the many differences will be highlighted in the
text. The most original feature here is the increase of the detected
asymmetries with the applied electric field. Our proposal
\cite{Ref11} to best take advantage of this feature, will be
presented briefly, keeping in mind the goal of APV measurements at
the 0.1$\%$ level.

\section{Principles of the experiment}\label{sec:3}
A pulsed, 539 nm laser beam excites the $6S_{1/2},F=3 \to
7S_{1/2},F=4$ HF transition of Cs while a longitudinal electric
field is applied, for a duration of only 100~ns not to trigger
discharges in the 8~cm long cell \cite{Ref12}. After this powerful
(1.5~mJ in 15~ns) excitation of the forbidden transition, a weak
(1~mW), infrared beam stimulates emission on the $7S_{1/2},F=4 \to
6P_{3/2},F=4$ allowed transition, for 20~ns. The pump and probe
beams are superposed with the same direction $\hat{k}$ and the same
linear polarisations $\hat \epsilon_{exc}\parallel \hat
\epsilon_{pr}$. The PV signal is the change of the probe
polarisation associated to the excitation of the forbidden
transition.
\\
An effective dipole operator \cite{Ref1} can be used to describe the
excitation of the forbidden transition:
\begin{equation}
-\hat \epsilon_{exc} \cdot \vec d_{eff} = \hat \epsilon_{exc} \cdot
(iImE_1^{pv}\vec\sigma - M_1^{'}\vec\sigma \wedge \hat{k}+ i\beta
\vec\sigma \wedge \vec{E} +\alpha \vec{E}).
\end{equation}
The first term is the PV electric dipole amplitude ($\vec\sigma$ is
the spin Pauli operator, and $iImE_1^{pv}$ is pure imaginary), the
second one is the magnetic dipole contribution, and the two last
ones are the Stark-induced, respectively vector and scalar,
contributions. To give orders of magnitude, the corresponding
partial $7S_{1/2}$ lifetimes are about 12~million years for the PV
electric dipole amplitude, 12 days for the magnetic dipole
amplitude, 300~s and 3~s, respectively, for the vector and scalar
Stark-induced transitions in a 2kV/cm $\vec E$ field. However, in a
longitudinal field $\vec E_{l}= E_l\hat{z}$, the $\alpha
\vec{E}\cdot\hat \epsilon_{exc}$ term is absent since the excitation
polarisation is transverse to $\hat{k}\parallel\hat{z}$. Also
important is the fact that, due to the choice of the linearly
polarised excitation, the real term $M_1^{'}\vec\sigma \wedge
\hat{k}$ is out of phase with the $i\beta \vec\sigma \wedge
\vec{E}_{l}$ pure imaginary term. This is a crucial difference with
the Boulder experiment, where these two terms interfere, which
contributes to several systematics \cite{Ref10}. The magnetic term
will be omitted throughout, except in sect.~\ref{sec:8} where it is
exploited for calibration.
\\
In the absence of the P violating term, the effective transition
operator (Eq.(1)) would be
%$-\hat \epsilon_{exc} \cdot \vec d_{eff}
$\hat \epsilon_{exc}\cdot (i\beta \vec\sigma \wedge
\vec{E}_{l})=-i\beta E_{l}\sigma_{x}$, assuming that $\hat
\epsilon_{exc}= \hat{y}$. Adding the PV contribution changes this
term into $-i\beta E_{l} \lbrace \sigma_{x}-(ImE_{1}^{pv}/\beta
E_{l})\sigma_{y}\rbrace $. Indeed, the PV term has the same,
geometrical consequence as a tilt of the excitation polarisation
\begin{equation}
\hat \epsilon_{exc} \to \hat \epsilon_{exc} + \theta^{pv}\hat{z}
\wedge \hat \epsilon_{exc}
\end{equation}
by an angle $\theta^{pv} = -ImE_1^{pv}/\beta E_l$.  As a result, the
eigen-axes of the excited state density matrix $\rho_{7S}\propto
(\beta E_{l}\sigma_{x})^2 \propto (\beta E_{l}F_{x})^2$ will be
tilted as well. Consequently, the vapour will act on the probe as an
anisotropic amplifier {\it with eigen-axes tilted}, by the angle
$\theta^{pv}$, {\it with respect to the symmetry planes} imposed by
the direction of the linear polarisation $\hat \epsilon_{exc}$. The
tilt angle $\theta^{pv}$ is small, $10^{-6}$ radian at $E_{l}
\approx$ 1.6~kV/cm, but is odd in $E_{l}$ reversal, a clear-cut
signature. Hence the PV signal will be a change of the linear
polarisation of the amplified probe beam, odd under $E_{l}$
reversal, manifesting the presence, in the probe gain matrix, of a
chiral contribution associated with the pseudoscalar ($\hat
\epsilon_{exc} \cdot \hat \epsilon_{pr})(\hat \epsilon_{exc} \wedge
\hat \epsilon_{pr} \cdot \vec E_l)$.
\\
Although the expected probe polarisation signal can be calculated
\cite{Ref13}, an efficient calibration has been obtained by
deliberately tilting  $\hat\epsilon_{exc}$ by a known angle
$\theta_{cal}$ and measuring the corresponding probe polarisation
signal in rigorously identical conditions. For $\theta_{cal}$ in the
milliradian range, proportionality is good enough \cite{Ref14} to
allow linear extraction of $\theta^{pv}$ from the PV signal. As
compared to the Boulder experiment, this calibration procedure is
obtained in real-time, and is free from any lineshape correction.

\section{Short review of the systematic effects}\label{sec:4}
Systematics can be classified as rank-1 or rank-2, according to the
number of defects involved.
\\
{\bf Rank-1}: {\it i)} A tilt of $\hat \epsilon_{exc}$, if it were
{\it odd} in the reversal of the applied electric field, would give
rise to a systematic effect: this is looked for and estimated by
real time monitoring of $\hat \epsilon_{exc}$. {\it ii)} Remembering
that a longitudinal magnetic field induces Larmor rotation of the
eigen-axes of the vapour, an $E_{l}$-odd $B_{z}$ field would give
rise to a systematic. Although the symmetry of the experiment
forbids even an $E_l$-even $B_{z}$ field, a significant fraction of
the data acquisition is devoted to the measurement of $B_{z}^{-}$,
using atomic signals (see sect.~\ref{sec:6}).
\\
{\bf Rank-2}: {\it i)} An offset between the pump and probe
polarisations, in the presence of an {\it imperfect reversal} of the
applied $\vec E_{l}$ field, could give rise to a systematic. However
the data processing (see sect.~\ref{sec:6}) is made in a way that
rejects their coupling. In addition, these two imperfections
monitored using specific signals are kept below $10^{-4}$ and
$10^{-3}$, respectively. {\it ii)} Defects that break cylindrical
symmetry. Ideally, simultaneous rotation of the pump and probe
polarisations (and the corresponding polarimeters) should leave the
experiment unchanged, except for possible {\it transverse} electric
or magnetic fields, or a misalignment between the excitation and
probe beam \cite{Ref15}. The coupling of a transverse electric field
to a transverse magnetic field gives rise, in the calibrated probe
polarisation signal, to a contribution
\begin{equation}
2\omega_{F^{'}}\tau \frac{E_{t}}{E_{l}} (\hat B_{t} \cdot \hat
E_{t}- (\hat B_{t}\cdot \hat \epsilon_{exc})(\hat E_{t}\cdot \hat
\epsilon_{exc})),
\end{equation}
in which $\omega_{F^{'}}\tau$ is the average Larmor precession under
the magnetic field $\vec B_{t}$ during the time spent by the atoms
in the excited state. This term mimics the PV $E_{l}$-odd term if
the $\vec B_{t}$ and $\vec E_{t}$ fields are {\it both} odd, or {\it
both} even, in $\vec E_{l}$ reversal. When averaged over two
polarisations at 90${^\circ}$ of each other, this systematic is
reduced by a factor of $1/2$ ("class 1" systematic effect). On the
other hand, the angular dependence, $ \hat B_{t}\cdot(\hat E_{l}
\wedge \hat \epsilon_{exc})\cdot(\hat B_{t}\cdot \hat
\epsilon_{exc})$, of the systematic due to the second-order
perturbation by a magnetic field completely cancels in the
90${^\circ}$ switch polarisation average: "class 2" systematic
effect. In ref. \cite{Ref15} can be found the inventory of both
class 1 (the most important ones) and class 2 systematic effects.

\section{Implementation of the experiment}\label{sec:5}
\begin{figure}
\resizebox{0.30\textwidth}{!}{%
\includegraphics{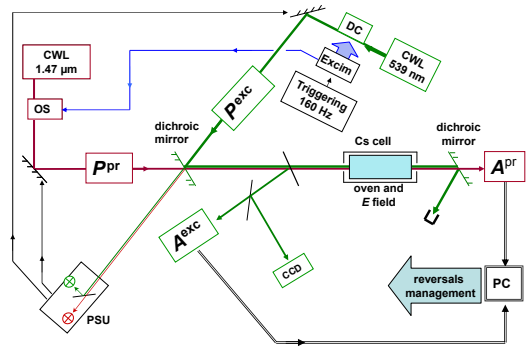}
}
\hspace{1.5cm}       % Give the correct figure height in cm
\caption{Main parts of the experimental set-up. CWL:
single-frequency, continuous wave laser. OS: sub-nanosecond optical
switch. P$_{pr}$, P$_{exc}$, A$_{pr}$, A$_{exc}$, polariser and
analyser units. Excim: excimer laser. DC: dye cells amplifier unit.
PSU: beam position stabilisation unit. CCD: camera for excitation
beam profile monitoring. PC: computer for data acquisition, and
management of the parameter reversals. The two-stage oven allows
seperate control of the Cs cell reservoir and wall temperatures.}
\label{fig:1}       % Give a unique label
\end{figure}
We briefly recall the experimental setup, more extensively described
in \cite{Ref16}, \cite{Ref17}, where the absolute frequency
stabilisation of the laser sources is presented.
\\
The excitation and probe beams are recombined using a dichroic
mirror (see Fig.~\ref{fig:1}). Beforehand, the polarisation of each
of the two beams has been prepared in a "polariser unit" made of a
calcite Glan polariser that defines the vertical polarisation, and
several half-wave plates that are mechanically inserted to work with
horizontal, or +/- 45${^\circ}$ polarisation. The excitation
polariser unit also includes {\it i)} a wave plate chosen to be
precisely $1\times \lambda_{exc}$ at {\it normal incidence}. When
appropriately tilted, it allows to compensate the birefringence in
the input Cs cell window; {\it ii)} a Faraday modulator, providing
the $\pm\theta_{cal}$ tilts required for the calibration procedure.
The probe polariser unit includes a tilted glass plate to compensate
for the linear dichroism in the recombination mirror. The "analyser
units" are based on a two-channel, balanced-mode polarimeter
\cite{Ref18}, preceded by several half-wave plates, two of which are
inserted synchronously with the half-wave plates of the
corresponding polariser unit, so as to operate the polarimeter
always in the same conditions. The third half-wave plate has its
axes at 45${^\circ}$ to the axes of the polarimeter. Inserting this
"cleaner" half wave plate allows to reject the polarimeter signals
that are not due to a beam polarisation change \cite{Ref19}. The
probe analyser records two kinds of signal: the first 20~ns probe
shot, immediately after excitation of the forbidden transition,
which contains the excited state contribution, and, one ms later, a
burst of four probe shots, which serves as a reference for the probe
polarisation signals.
\\
Sapphire appeared as an appropriate material for the cell body, due
to its very low surface conductivity in the presence of caesium
vapour \cite{Ref20}. This should have suppressed the $B_{z}^{-}$
field. However, a considerable electron charge (and corresponding
current) was observed at the anode window of the sapphire cells,
until we could mechanically suppress the multiplication, due to
secondary emission, of the electrons emitted at the cathode window
\cite{Ref21}. Then the $B_{z}^{-}$ field was observed to be at most
a few tens of $\mu$G, corresponding to a false effect of a fraction
of the PV signal (11$\%$ on average), for which we can correct
practically in real time. Measurement of $B_{z}^{-}$ uses the
(large) optical rotation signal on the F=4 to F=5 probe transition
\cite{Ref22}.

\section{Measurement procedure}\label{sec:6}
The basic element of the data acquisition is the simultaneous
recording of the two signals of the two-channel probe polarimeter,
to form the imbalance $D^{ampl} \equiv (S_1 - S_2)/(S_1 + S_2)$ with
the {\it amplified} probe beam. After the reference probe pulses
have been detected, the corresponding imbalance is subtracted:
$D_{at} \equiv D^{ampl} - D^{ref}$, and this is repeated for 30
shots (duration 200~ms) before $\theta_{cal}$ is reversed. The next
reversal is that of the electric field (period 800~ms). Insertion
and removal of the "cleaner" half-wave plate is repeated every 7~s.
The input probe polarisation undergoes a 90$^\circ$ switch every
14~s. A value $\theta^{pv}$ is then obtained by $\lbrack
\theta^{pv}\rbrack _{\hat \epsilon_{exc}}= \theta_{cal}\times$
\begin{equation}
\left < \sigma_{_E} \left [\frac{ < \sigma_{clean} D_{at} (\lbrace
\sigma_j  \rbrace) >_{\sigma_{clean} \sigma_{cal}}}{< \sigma_{clean}
\sigma_{cal}D_{at}(\lbrace \sigma_j \rbrace ) > _{\sigma_{clean}
\sigma_{cal}}} \right ]   \right >_{ \sigma_{_E}
\sigma_{90^{\circ}pr}}
\end{equation}
where $\sigma_{i}$ indicates the state of reversal (or plate
insertion) {\it i}. Beside this parity violating, linear dichroism,
other linear combinations are monitored for diagnosis of possible
drifts or systematics. The probe optical rotation, obtained by
inserting a factor $\sigma_{90^\circ pr}$ after the $\sigma_{E}$
factor in equation above, is used for measuring the $B_{z}^{-}$
field. In order to suppress systematics (see sect.~\ref{sec:4}), two
"isotropic values" of $\theta^{pv}$ are obtained, every 5~mn, after
$\hat \epsilon_{exc}$ has been switched by 90$^\circ$, then $\pm
45^\circ$:
\begin{equation}
S_{xy} \equiv \frac {1}{2} (\lbrack \theta^{pv} \rbrack_x + \lbrack
\theta^{pv} \rbrack_y){\it~and~} S_{uv} \equiv \frac {1}{2} (\lbrack
\theta^{pv} \rbrack_u + \lbrack \theta^{pv} \rbrack_v).
\end{equation}
The recording of isotropic values goes on for about 90~mn, after
which a last reversal is performed, concerning the cell orientation
with respect to the light beams. Operating at normal incidence at
the cell windows gives rise to an excess noise on the polarimetric
measurement, due to an etalon, interference effect between the cell
windows and optical surfaces inside the polarimeter. To prevent
this, the cell is tilted by an angle $\psi = 3\times 10^{-3}$~rad.
However, the atoms excited by the 539~nm beam reflected at the exit
cell also contribute. This beam is tilted by an angle 2$\psi$ with
respect to the probe beam, and it can be shown that the rank-2
systematic associated with this tilt coupled to a transverse
electric field is linear in $\psi$ \cite{Ref15},\cite{Ref17}. It is
suppressed by  {\it i)} reversing the angle $\psi$, and {\it ii)}
reducing the window reflection (see sect.~\ref{sec:7}).
\\
Finally, a PV data-taking run starts and closes with the measurement
of the optical rotation on the 4$\to$5 probe transition to determine
the value of the stray $B_{z}^{-}$ field and correct for the
associated contribution. This lengthens data acquisition by about
60$\%$. On the other hand, the measurement of the stray, transverse
electric and magnetic fields, made by application of large,
controlled, transverse magnetic fields, does not increase
significantly data taking duration, although it is regularly
repeated throughout the data acquisition. In the results presented
below, the systematics are controlled  and kept at the 1$\%$ level.
The 2.6$\%$ total error bar on the average is mainly statistical,
and includes the slight contribution due to the $B_{z}^{-}$
systematic correction.

\section{The PV results and their improvement}\label{sec:7}
\begin{figure}
%\resizebox{0.30\textwidth}{!}{%
%\includegraphics{figure2.eps}
%}
\centerline{\epsfxsize=50mm \epsfbox{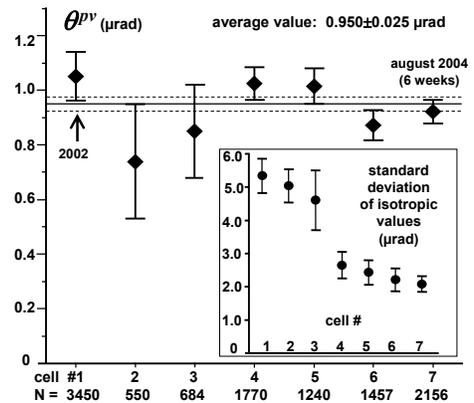}}
%\vspace{0cm}       % Give the correct figure height in cm
\caption{The values of $\theta^{pv}$ measured in the seven
successive cells, at electric field 1.62kV/cm. Inset: the standard
deviation, as a function of cell number. Adapted from
Ref.\cite{Ref17} with permission from APS.}
\label{fig:2}       % Give a unique label
\end{figure}
Seven cells have been used successively for the measurements. The
measured values of $\theta^{pv}$ are displayed in Fig.~\ref{fig:2}.
For each cell the number N of isotropic values recorded is
indicated. The first measurement took place in 2002 \cite{Ref24} and
confirmed the validity of the method. Later, efforts devoted to the
reduction of the noise and of the systematics have improved the
statistics, as is obvious from the inset of Fig.\ref{fig:2}: the
standard deviation of the isotropic values recorded in cell $\#7$ is
2.5 times smaller that in cell $\#1$.
\\
This has resulted from implementing:
\\
- sapphire cells with highly parallel windows \cite{Ref25}. The
fraction of the excitation beam reflected at each window can be made
very small ($10^{-3}$ instead of 5$\%$ in cell $\#1$) by choosing a
temperature such that the reflection is cancelled by interference.
This further suppresses the systematic effect associated with the
tilted reflected beam (see sect.~\ref{sec:6}). It also suppresses
the interference between the beams reflected at the input and exit
windows.
\\
- a "polarisation magnifier" \cite{Ref26}. This device, made of 4 or
6 plates at Brewster incidence, enhances the angle of the
polarisation tilt to be measured by the polarimeter. Since it
attenuates the beam, it allows higher probe power without changing
the polarimeter preamplifier chain, improving the signal/noise ratio
when the photon shot noise contribution is significant.
\\
- a better extinction for the probe optical switch. The probe
intensity that leaks through the closed optical switch can
contribute to the noise.
\\
- long term absolute frequency stabilisation of the excitation
\cite{Ref17}.
\\
A statistical analysis \cite{Ref17} shows that the values obtained
in the different cells agree. Indeed the cells were not identical:
the material of the cell tube (alumina ceramics {\it vs}
monocrystalline sapphire), the polishing/origin/tilt of the windows,
the filling of the cells (unexpected foreign gas observed in cell
$\#$3) etc. were different. The agreement brings confidence in the
absence of a bias from cell imperfections. The fact that the
averages of the $S_{xy}$ and  $S_{uv}$ values agree to about 0.7$\%$
also brings confidence as to the cylindrical symmetry of the
experiment.

\section{Calibration aspects; measurement of the  $\vec E_l$ field}\label{sec:8}
First, the measurements presented above rely on the determination of
the Faraday rotation angle used for calibration of the PV linear
dichroism signal. This was done by different methods, in particular
by measuring the mechanical rotation of the polariser that
compensates for the Faraday rotation, with an accuracy of about
0.5$\%$, that could be improved if need be.
\\
Second, the determination of $\theta^{pv}=-ImE_1^{pv}/\beta E_l$ can
be exploited only if $E_l$ is known with enough accuracy {\it inside
the vapour}, {\it i.e.}, measured by the atoms themselves. This can
be done by measuring the Stark parity conserving alignment,
proportional to $\theta_{cal} \beta^{2} E_l^{2}$, and then measuring
the zero-field alignment, proportional to $M_1^{'2}$. The
proportionality factors are the same, except for the influence of
saturation by the probe beam: in the presence of the applied,
$\approx$~1.6kV/cm electric field, the amplification of the probe
beam is non-negligible, while it is very small, and buried in noise
and background, in the absence of electric field. An accurate
calibration of the electric field can be obtained by the comparison
of these two, similar, linear dichroism signals, both measured {\it
vs} probe beam intensity $I_{pr}$ and extrapolated to $I_{pr}=0$:
$%\begin{equation}
E_l^{exp}=  \frac{M^{'}_1}{\beta} \left (\sqrt{ \frac{\ln{\lbrack 1+
D_{at}(E=E_l,\theta=\pm\theta_{cal} )/2\theta_{cal}} \rbrack }{(1+
\epsilon) D_{at}(E=0, \theta= \pm\pi/4)}}-1 \right) $%\end{equation}
\\
Here the small $\epsilon$ quantity expresses the deviation of the
exact result, calculated numerically \cite{Ref13}, with respect to
the simple one assuming an exponential-type amplification. In our
experimental conditions \cite{Ref17}, we obtain $\epsilon=0.100$,
with a resulting uncertainty of 0.3$\%$ on $E_l^{exp}$. Presently
the uncertainty in $E_l^{exp}$ is negligible as compared to the
$2.5\%$ uncertainty in the determination of $\theta^{pv}$.
\section{Conclusions and perspectives}\label{sec:9}
For the measured $E_l^{exp}=1.619$~kV/cm we obtain
$\theta_{exp}^{pv}=(0.950\pm0.025)~\mu rad$, in excellent agreement
with the value of $0.962\pm0.005$ obtained from the Boulder
measurement \cite{Ref9} on the same 6S,~F=3~$\to$~7S,~F=4 hyperfine
transition. Using the value of the $\beta$ polarisability published
in \cite{Ref27}, the corresponding value of the P violating
transition dipole, $ImE_1^{pv}(3\to4)=-0.808 \times 10^{-11} ~\vert
e \vert a_0$, is obtained with a $2.1\times 10^{-13}~\vert e \vert
a_0$ absolute uncertainty.
\\
Our experimental method is very different from that used by the
Boulder group \cite{Ref10}, with many consequences as regards
calibration, or systematics. It makes use of the amplification of a
probe beam, the higher the gain, the larger the asymmetry to be
measured. Hence, the PV asymmetry is an {\it increasing} function of
the applied electric field. One important consequence is the
considerable enhancement of sensitivity that can be foreseen in an
experiment with a {\it transverse} electric field and a longer
interaction length \cite{Ref11}. Despite the transverse field
configuration, a special multi-electrode design can "restore
cylindrical symmetry", which helps in tracking and rejecting
systematic effects \cite{Ref28}. With reasonably moderate values of
the probe gain (still far from spontaneous superradiance), a
sensitivity of 0.1$\%$ looks a realistic objective, although the
regime of higher gains ({\it triggered} superradiance) probably
deserves specific attention. The suggested experiment involves the
excitation in the presence of transverse magnetic and electric
fields, with two counter-propagating excitation beams. It is further
motivated by the recent progress in the atomic physics calculations
of the factor that links the nuclear spin independant value of
$E_1^{pv}$ to the Cs nucleus weak charge, now aiming at the 0.1$\%$
accuracy \cite{Ref29}.

\end{document}